# Effect of in-situ electric field assisted growth on anti-phase boundaries in epitaxial $Fe_3O_4$ thin films on MgO

Ankit Kumar[1]*, Erik Wetterskog[1], Erik Lewin[2], Cheuk-Wai Tai[3], Serkan Akansel[1], Sajid Husain[4], Tomas Edvinsson[1], Rimantas Brucas[1], Sujeet Chaudhary[4], and Peter Svedlindh[1#]

[1]Department of Engineering Sciences, Uppsala University, Box 534, SE-751 21 Uppsala, Sweden
[2]Department of Chemistry - Ångström Laboratory, Uppsala University, Uppsala, Sweden
[3]Department of Materials and Environmental Chemistry, Stockholm University, Stockholm Sweden
[4]Department of Physics, Indian Institute of Technology Delhi, New Delhi, India

Anti-phase boundaries (APBs) normally form as a consequence of the initial growth conditions in all spinel ferrite thin films. The presence of APBs in epitaxial films of the inverse spinel $Fe_3O_4$ alters their electronic and magnetic properties due to strong antiferromagnetic (AF) interactions across these boundaries. The effect of using in-situ electric field assisted growth on the migration of APBs in hetero epitaxial $Fe_3O_4(100)$/MgO(100) thin films have been explored in the present work. The electric field assisted growth is found to reduce the AF interactions across APBs and as a consequence APBs free thin film like properties are obtained, which have been probed by electronic, magnetic and structural characterization. An increase in energy associated with the nucleation and/or early stage of the growth and, therefore, a corresponding increase in surface mobility of the ad-atoms play a critical role in controlling the density of APBs. This innovative technique can be employed to grow epitaxial spinel thin films with controlled AF interactions across APBs.

* E-mail: ankit.kumar@angstrom.uu.se

# E-mail: peter.svedlindh@angstrom.uu.se

## I. INTRODUCTION

Magnetite ($Fe_3O_4$) is one of the oldest known magnetic materials with a wide range of applications; as magneto receptors used by certain organisms to perceive direction[1], as catalyst in surface science[1], or as magnetic labels in bioassays[1]. In recent years, magnetite has acquired increased interest from researchers not only due to being a strongly correlated electron system but also due to being an attractive material for the nascent field of spintronics because of its half-metallic character, high conductivity ($\sim 2 \times 10^{-4}\ \Omega^{-1} cm^{-1}$), and its high Curie temperature ($T_C = 858\ K$).[2,3,4,5] One of the most interesting physical phenomena associated with $Fe_3O_4$ is the occurrence of a metal-insulator (MI) transition at ~120K (VT), the so-called Verwey transition (VT), revealed by the increase in resistivity and the sudden drop in magnetization.[4,5] The VT is very sensitive to stoichiometry variations, stress, and magnetic disorder in the $Fe_3O_4$ system (more details about the VT can be found in Ref. 5). Despite having profound magneto-electronics compatible properties, $Fe_3O_4$ thin films have so far not been employed to their full potential. A snag with $Fe_3O_4$ thin films is that during film growth ionic disorder at the tetrahedral and octahedral sites in the crystal structure gives rise to antiphase boundaries (APBs).[6] The presence of APBs degrades both the electronic and magnetic properties, e.g. by reducing the half-metallicity, by increasing the resistivity and by delaying the approach to saturation due to strong antiferromagnetic (AF) interaction across boundaries.[7,8,9] The APBs are abundant in epitaxial $Fe_3O_4$ thin films grown on MgO substrates, and the strength of the associated AF interaction depends on the angle between neighbouring anti-phase domains.[6,7,8,9,10] Eerenstein et al. reported that APBs are diffusive in nature,[11] since the free energy of APB dominated thin films is higher than for APB-free films, making them unstable. Consequently, annealing of these films results in migration of APBs. By increasing the energy associated with the nucleation and/or early stage of the growth there will be a corresponding increase in surface mobility of the ad-atoms that can play a critical role in the reduction of APBs.[8,9,12] High temperature growth has the potential to reduce their density.[8] However, high temperature growth may result in inter-diffusion at the film-substrate interface which lowers the mobility of coalescing ad-atoms/clusters, and hence can also lead to a high density of APBs. Since MgO is one of the outstanding barrier materials in magnetic tunnelling junctions, growth of half-metallic (100 % spin polarization) thin films over barrier materials such as MgO materials is a key ingredient in spintronic devices. The lattice mismatch between MgO (a = 4.212 Å) and $Fe_3O_4$ (a = 8.396 Å) is ~ 0.33% and has been used extensively to grow epitaxial films by various deposition techniques,[6,7,11,13,14,15,16] however, without significantly reducing the density of APBs in the films.

Pulsed dc sputtering involves low energy sputtered species, and is not expected to provide significant ad-atom mobility, like other deposition methods, thereby not favouring the migration of APBs. However, by using an in-situ electric field applied during film deposition sufficient energy may be provided for transportation of ad-atoms towards the growing nuclei on the substrate surface and may therefore result in migration and reduction APBs that forms during coalescence of nuclei in the growing film. In this paper, we report the effect of using an in-situ electric field during low temperature growth of epitaxial $Fe_3O_4$(100) thin films on MgO(100) substrates on the density of APBs. The in-situ electric field results in a rearrangement of APBs along with the formation of nanocrystallinity, in such a way that boundaries with strong AF coupling which yield a delayed saturation of the magnetization and reduced remanent magnetization are suppressed.

## II. EXPERIMENTAL METHODS

Epitaxial $Fe_3O_4$ thin films were grown on MgO(100) substrates using pulsed dc reactive sputtering. Prior to $Fe_3O_4$ deposition the MgO surface was cleaned by vacuum annealing and monitored by RHEED. The $Fe_3O_4$ phase purity was optimized by varying the oxygen partial pressure from 1.5 to 3.5 $\times 10^{-5}$ Torr in steps of 0.5 $\times 10^{-5}$ Torr, keeping constant substrate temperature, $T_S$ = 300 °C, and sputter power (85W). In contrast to the conventional method to grow phase pure epitaxial $Fe_3O_4$ on MgO we have employed a low substrate temperature ($T_S$) to prevent the interface diffusion which is quite prominent around and above $T_S$ = 400°C. The film deposited at 2 $\times 10^{-5}$ Torr oxygen partial pressure exhibits the pure $Fe_3O_4$ phase, and is referred to as the as-deposited (ASD) sample. The thickness of the ASD film was 35 nm. During the process of in-situ electric field assisted growth optimization, different voltages (1V, 10V and 100 V) were applied between two pre-sputtered gold contacts on the MgO(100) substrate with a gap of 6 mm. Optimized conditions were obtained for 10V potential difference, which was chosen for a detailed study where another set of $Fe_3O_4$ films was deposited in the presence of the in-situ electric field. The other growth parameters were kept the same as for the ASD sample. It is important to note that the in-situ electric field did not change the temperature of the growing film. Figure 1 shows a schematic of the setup used for in-situ electric field assisted growth of epitaxial $Fe_3O_4$ thin films in a ultra-high vacuum chamber equipped with a differentially pumped high pressure RHEED system.[8] Two electric field assisted thin films with thickness 20 nm and 30 nm were deposited, named hereafter as ED1 (20 nm) and ED2 (30 nm), respectively.

X-ray diffraction $\theta/2\theta$ and pole figure measurements were performed employing a PANalytical MRD instrument using a Cu-$K_\alpha$ radiation source and parallel beam optics. The measurements were conducted using a parallel plate collimator with a 0.18° acceptance angle, and a flat graphite monochromator on the secondary side (to supress the fluorescence). In the $\theta/2\theta$ measurements line focus and a Goebel mirror were used on the primary side, and pole figures were acquired using point focus and a poly-capillary X-ray lens on the primary side. The pole figures were measured using the $Fe_3O_4(311)$ reflection at $2\theta = 35.42°$, which also detects the tail of the MgO(111) reflection centred at $2\theta = 36.95°$.

Bright- and dark-field TEM images were taken using a JEOL JEM-2100F field emission gun microscope (point resolution = 1.9Å, $C_s$=0.5mm, $C_c$=1.1mm) equipped with a Gatan Ultrascan 1000 and a Orius 200D camera. The samples were prepared by dissolving the MgO substrate in 6 wt% ammonium sulphate solution (72 h, at 20 °C), thus under slightly different conditions as those described in Eerenstein et al. [11,13] After dissolution of the MgO, the thin films were placed on carbon mesh films.

A Renishaw InVia micro Raman set-up using a 532 nm laser and 1800 lines/mm grating was used to record the Raman spectra on all the samples at 2 mW power. Lower laser intensities down to 100 $\mu$W were also used to double-check that no laser induced changes of the samples occurred. Isothermal magnetizations (*M*) versus field (*H*) as well as magnetization versus temperature (*T*) measurements were performed on all samples using a Quantum Design MPMS system.

III. RESULT AND DISCUSSION

To confirm epitaxial growth and crystallographic orientation relationships of the $Fe_3O_4$ thin films deposited on MgO(100) substrates X-ray diffraction $\theta/2\theta$ and pole figure measurements were performed. X-ray diffractograms are shown in Fig. 2 for all samples, revealing two strong peaks situated at $2\theta = 42.9°$ and $94.0°$, respectively. These two Bragg peaks correspond to (004) and (008) Fe$_3$O$_4$ reflections but also agree well with the (002) and (004) Bragg reflections of the MgO substrate.[11,15,17,18]

The contributions from the substrate and the deposited film are thus overlapping and the exact determination of the lattice parameter is therefore difficult by $\theta/2\theta$ analysis. However, the data clearly indicates a strained epitaxial cube-on-cube growth of the Fe$_3$O$_4$ thin film which is expected since the lattice mismatch between Fe$_3$O$_4$ and MgO is ~0.33 % yielding a small but finite tensile strain. Conversely, it is important to mention here that presence of APBs in epitaxial Fe$_3$O$_4$(100)/MgO(100) results in compressive stress (more details are given in Ref. 14 and references therein).[11,14,19] Furthermore, the ED1 and ED2 peak intensities are a bit smaller which could be due to smaller film thickness. The ED1 and ED2 films exhibit an additional weak signature at 65.2° corresponding to the Fe$_3$O$_4$(531) reflection, but the concentration of crystallites with this particular orientation is roughly three orders lower than that of crystallites with the (400/800) orientations parallel to the substrate. The presence of this minute misorientation and the formation of nanocrystallinity (see Fig. 4e) in ED films can be linked to the electric field assisted growth.[14] Since the presence of APBs in epitaxial Fe$_3$O$_4$(100)/MgO(100) yields compressive strain, the in-situ electric field assisted migration/rearrangement of APBs results in strain relaxation,[14] which in turn results in the observed misorientation in ED films.

To further confirm the epitaxial and/or orientation relationship pole figure analyses were performed using the Fe$_3$O$_4$(311) reflection, as shown in Figs. 3(a), 3(b), and 3(c) for the ASD, ED1 and ED2 films, respectively. As inferred from the pole figures, all films exhibit similar patterns that correspond to the epitaxial relationship Fe$_3$O$_4$ (001)[110] // MgO (001)[110], i.e., cube-on-cube growth. Both peaks at the out of plane tilt angles $\psi$ = 25° and 73° originate from the Fe$_3$O$_4$(001) orientation. The narrow peaks at $\psi$ = 55° originate from the substrate and match the angle between the MgO(002) and MgO(111) reflections, of which the tail of the latter is detected in the present pole figure measurements, thus confirming the in-plane relationship between the substrate and film. The minority orientation corresponding to Fe$_3$O$_4$(531) observed for the electric field assisted deposited films is not detected in the pole figures, thus implying that the misorientation exists only in minute amount.

Figure 4a shows the dark-field transmission electron microscope (DFTEM) image of the ASD sample. Imaging the sample using the weak 220 reflection (see selected area electron diffraction (SAED) pattern in the inset) reveals the APBs in the single-crystalline Fe$_3$O$_4$ thin film, with a domain size comparable to those observed in previous studies.[6,11,13] In contrast, the bright-field image of the ED2 sample shown in Fig. 4b reveals a nanocrystalline microstructure, albeit with a strongly textured electron diffraction pattern (see Fig. 4c). The contrast in the ED2 DFTEM image shown in Fig. 4d, displays diffraction contrast due to small misalignment of the nanocrystal grains, but lacks the characteristic APB zig-zig pattern seen in Fig. 4a. Indeed, at higher magnifications (Fig. 4e), Moiré-fringes are clearly visible due to overlap between grains with similar orientation; thus, although the film is thin and strongly textured, it cannot be considered columnar.

To further confirm the phase purity, presence of APBs and strain in the samples unpolarised micro Raman measurements were performed. The Raman spectra were recorded in the 100 to 800 cm$^{-1}$ frequency range, and only Fe$_3$O$_4$ $A_{1g}$ and $T_{2g}$ Raman modes were observed. The recorded $A_{1g}$ and $T_{2g}(3)$ modes for all films are shown in Fig. 5. The observed Raman spectra were fitted with a sum of Lorentzian functions, shown in Fig. 5, to extract the line shape parameters; angular frequency ($\omega$) and full width at half maximum linewidth FWHM ($\Gamma$). The observed values of $\omega$ and $\Gamma$ are shown in Table I. The errors in obtaining $\Gamma$ and $\omega$ are about 0.5 and 0.5 cm$^{-1}$ for the $A_{1g}$ mode and 1.5 and 0.5 cm$^{-1}$ for the $T_{2g}(3)$ mode, respectively. The observed $\omega$-values for the $A_{1g}$ and $T_{2g}(3)$ modes of the epitaxial films are lower than the corresponding Fe$_3$O$_4$ single crystal values; $\omega$ = 669 and 307 cm$^{-1}$ for the $A_{1g}$ and $T_{2g}(3)$ modes, respectively.[4,12,20,21] The ASD(35nm) and ED2(30nm) films exhibit ~0.21% and ~0.12 % tensile strain, respectively, whereas the ED1(20nm) film possesses ~0.45% tensile strain. Here it should be stressed that the strain is the cumulative effect of the APB compressive strain, and the Fe$_3$O$_4$ and MgO lattice mismatch induced tensile strain. It is well known that hetero-epitaxial growth of Fe$_3$O$_4$ on MgO is strained and that this strain is known to be accommodated until 70nm film thickness; a thickness much above the critical thickness predicted by the theoretical model for strain relaxation.[14,19] The compressive strain generated by APBs and the lattice-mismatch induced tensile strain compensate each other to some extent and therefore allow the ASD film to grow mostly coherently with the substrate[14], which is evident from pole figure and Raman results, in accordance with previous reports.[14,19] However, the presence of the electric field during growth provides extra energy to nuclei/ad-atoms in the nucleation process of the film growth and helps to migrate the APBs that form during the nucleation process.[8,9,12]

To explore the presence of APBs and/or AF interactions in the samples Allen's approach[22] was employed. Originally Allen showed that the electron-phonon (*e-p*) coupling in disordered metallic systems could be estimated by the wave vector averaged phonon linewidth,[22] since the conservation rule relaxes in the presence of disorder among different wave vectors leading to the decay of phonons into electron-hole pairs, and as a result the *e-p* interaction is enhanced.[4] This approach provides an estimation of the degree of disorder in the system by enhancing the e-p interaction. In epitaxial $Fe_3O_4$ thin films, ionic disorder results in the formation of the APBs and therefore in the enhancement of the e-p interactions. Moreover, the nanocrystallinity along with the small misorientation found for ED1 and ED2 can further increase the value of the *e-p* interaction. The electron-phonon coupling strength can thus be an indirect indicator of the increase in density of APBs as well as of AF interactions across grain boundaries.[4] The strength of the *e-p* coupling according to Allen's approach can be expressed for the $i^{th}$ phonon mode as,

$$\lambda = \frac{\Gamma_i}{\omega^2} \frac{g_i}{2\pi N(E)}, \qquad (1)$$

where $g_i$ is the degeneracy of the $i^{th}$ mode, $g_i(T_{2g}) = 3$, $g_i(A_{1g}) = 1$, and $N(E)$ (=3 states/eV per Fe at room temperature) is the density of states (DOS) at the Fermi level[4,8]. The estimated values of the *e-p* coupling strength of the $A_{1g}$ and $T_{2g}(3)$ Raman modes are given in Table I for all the samples. In $Fe_3O_4$, the $A_{1g}$ mode which is due to symmetric stretching of the oxygen atoms along the Fe-O bonds is mainly connected to structural properties, while the $T_{2g}$ mode which is due to symmetric and asymmetric bending of oxygen with respect to Fe is linked to the electronic properties[4,23]. Therefore, growth induced changes in APBs are linked to the changes in the *e-p* coupling strength of the $T_{2g}$ mode. The ASD film exhibits a lower value of $\lambda \approx 0.68 \pm 0.02$ compared to the value $\lambda \approx 0.75 \pm 0.02$ obtained for the ED1 and ED2 films. Nevertheless, the observed values of $\lambda$ for films studied here are still quite low compared to previously reported values (~0.92 or greater) for epitaxial thin films of $Fe_3O_4$ on MgO,[24] and comparable to $\lambda$ values reported for significantly reduced density of APBs in two dimensional epitaxial $Fe_3O_4$ films grown on TiN buffered Si(100).[8,12] The DFTEM image (Fig. 4e) reveals the absence of APBs in the ED films. Hence, the value of the e-p coupling constant is expected to be similar to that of bulk single crystalline $Fe_3O_4$. Therefore, the small increase in $\lambda$ for the electric field deposited films compared to the ASD film can be explained by disorder at the nanocrystal grain boundaries. Here it is important to note that in our previous work[25] the in-situ electric field was applied to the polycrystalline growth of $Fe_3O_4$ on naturally oxidized Si(100) substrate and as a result the electric field assisted grown film exhibited preferred oriented growth in conjunction with a significantly reduced density of APBs. The present study is different as here the in-situ electric field was applied during epitaxial growth of $Fe_3O_4$ on MgO, and APBs are typically more abundant in epitaxial films of $Fe_3O_4$ grown on MgO substrates.

To further investigate the effects of in-situ electric field assisted growth the results from magnetization measurements will be explored. The APBs cause AF couplings in polycrystalline (inter- and intra-grain) and epitaxial (intra-grain) films by superexchange interactions at the cationic/anionic sites.[9,25] The strength of the AF interaction depends on the relative orientation of neighbouring crystallites.[6,11] Strong AF superexchange interactions at APBs exist between $Fe^{3+}$ and $Fe^{2+}$ ions at octahedral and $Fe^{3+}$ ions at tetrahedral sites via $O^{2-}$, having 180°, 140° and 120° angles for octahedral-$O^{2-}$-octahedral, tetrahedral-$O^{2-}$-tetrahedral, and octahedral-$O^{2-}$-tetrahedral couplings, respectively (for more details, see Ref. 6). The presence of these interactions inevitably results in reduced remanent magnetization ($M_r$) and enlarged saturation magnetic fields ($H_S$), as reported previously for epitaxial growth of $Fe_3O_4$ thin films over MgO substrates.[7,11,13,14,15,16] However, interactions across APBs can also be weakly AF or even ferromagnetic (FM) in nature.[6] Magnetizations versus field measurements have been performed at different temperatures on all samples; the results are shown in Figs. 6(a)-(c). The inset in each figure shows the temperature dependence of the saturation magnetization ($M_S$) estimated applying an in-plane magnetic field of 400 mT. To be able to compare $M_r$ and $H_S$ values for ASD and ED films, normalized in-plane magnetization loops measured at 300 K are shown in Fig. 6(d). The values of $H_S$ ($M_r/M_S$) for the ASD, ED1 and ED2 films are found to be 300 (61%), 150 (87%) and 150 (90%) mT, respectively. The observed values of $\mu_0 H_S$ ($M_r/M_S$) are quite small (large) compared to values of $\mu_0 H_S =$ 400 mT ($M_r/M_S = 70$ %) reported in our previous work wherein the in-situ electric field was applied during the polycrystalline growth of $Fe_3O_4$ on naturally oxidized Si(100).[25]

Epitaxial $Fe_3O_4(110)$ and $Fe_3O_4(111)$ films typically exhibit larger in-plane four-fold magnetic anisotropy compared to epitaxial $Fe_3O_4(100)$ films, and presence of APBs further supresses the signature of four-fold magnetic anisotropy in $Fe_3O_4(100)$ films.[12] The ASD film exhibits minute or insignificant in-plane four-fold magnetic anisotropy (results not shown). However, the electric field assisted deposited films exhibit clear evidence of four-fold magnetic anisotropy as shown in Fig. 6(e) for the ED1 sample. The cubic anisotropy and the large remanent magnetization in our ED films are comparable to that obtained in $Fe_3O_4$ single crystals [Ref. 12 and references therein]. The low $H_S$ and high $M_r$ in the ED films as compared to the ASD film clearly affirm the reduced number of APBs with strong AF exchange interactions in electric field assisted grown films. The applied in-situ electric field promotes the migration and rearrangement of APBs. In particular our results show that APBs with strong, high energy AF interaction across the boundaries will migrate and rearrange to form weak, low energy AF interactions.[6,10,11] As a result of this migration, comparing with the results obtained for the ASD film, the $M_r$ increases while $H_S$ decreases. It is worth mentioning that the in-plane saturation field of the ASD film itself is quite low compared to the values reported in literature; $\geq 1000$ mT.[7, 11, 13,14,15,16]

The comparison of in-plane and out-of-plane magnetization loops for the ED1 and ASD films (Fig. 6(f)) reveals something unexpected (similar results are obtained for the ED2 film); the out-of-plane saturation field ($H_{S,\perp}$) for the ED1 film is much larger than the expected value of $H_{S,\perp} \sim M_s$. It can therefore be concluded that the ED films exhibit a strong contribution to the magnetic anisotropy favouring in-plane magnetization. To further look into the effects of electric field assisted growth, field-cooled magnetization versus temperature results for the ASD and ED2 samples having comparable thickness are plotted in Fig. 7. The presence of magnetic disorder due to strong AF interactions across APBs and strain in the film can broaden and shift the Verwey transition ($T_V \sim$ 120 K for single crystals).[8,19] The Verwey transition for the ED2 sample appears approximately at 119 K, while the Verwey transition for the ASD film is in comparison sharp but shifted to lower temperature ($T_V \sim$ 112 K), presumably due to the presence of magnetic disorder created by strong AF interactions across APBs.[19] The Verwey transition temperature of the ED2 film is nearly equal to the bulk value, whereas $T_V \sim$ 112 K of ASD is lower due to presence of strong AF couplings across APBs. However, the drop in the magnetization at the Verwey transition is small in the ED2 film compared to the ASD film. A tentative explanation for the absence of a sharp change in magnetisation at $T_V$ for electric field assisted grown films is the presence of large in-plane anisotropy as evidenced by the large value of the out-of-plane saturation magnetic field.

## IV. CONCLUSION

Structural and magnetic properties of epitaxial $Fe_3O_4$ thin films deposited on MgO substrates by using in-situ electric field assisted growth have been compared with the corresponding properties of as-deposited films. The comparison emphasizes the importance of migration and rearrangement of APBs to reduce the strength AF interactions across these boundaries. Both ASD and ED films grow epitaxially, ED films with minor crystallographic misorientation, with the orientational relationship $Fe_3O_4$ (001)[110] // MgO (001)[110]. The important effects of in-situ electric field assisted growth were revealed by in-plane and out-of-plane magnetization versus field measurements. Both ED1 and ED2 exhibit a large remanent magnetization ($M_r/M_S \approx$ 90 %) and small saturation field ($\mu_0 H_S \approx 150\ mT$), clearly showing that the AF interactions across APBs have been greatly reduced in these films as also indirectly evidenced by the TEM dark field images (cf. Fig. 4e). Moreover, the enhanced out-of-plane saturation fields ($H_{S,\perp}$) observed for ED1 and ED2 show that the electric field induces a strong in-plane anisotropy. We would like to conclude that this innovative in-situ electric field assisted growth technique will be an excellent tool for the manipulation of magnetic properties in spinel and similar systems being of importance for magneto-electronic device technology.

## ACKNOWLEDGEMENTS

This work is supported by the Knut and Alice Wallenberg (KAW) Foundation Grant No. KAW 2012.0031. The KAW Foundation is acknowledged for an equipment grant for the electron microscopy facilities at Stockholm University.

**Table I** The Raman line shape parameters and corresponding electron phonon coupling strength for the different $Fe_3O_4$ film samples.

| Mode | $A_{1g}$ | | | $T_{2g}(3)$ | | |
|---|---|---|---|---|---|---|
| | ASD | ED1 | ED2 | ASD | ED1 | ED2 |
| $\omega(cm^{-1})$ | 667.6 | 663.7 | 668.2 | 306.7 | 304.7 | 307.2 |
| $\Gamma\ (cm^{-1})$ | 37.5 | 39.1 | 37.9 | 50.3 | 53.9 | 55.2 |
| $\lambda$ | 0.036 | 0.038 | 0.037 | 0.68 | 0.74 | 0.75 |
| $\lambda$ = 0.51 for bulk single crystals, and 0.72 for APBs free epitaxial thin films.[4,8] | | | | | | |

**Figure 1.**

**Figure 1** Schematic layout of in-situ electric field assisted pulsed dc sputter deposition of $Fe_3O_4$ thin films.

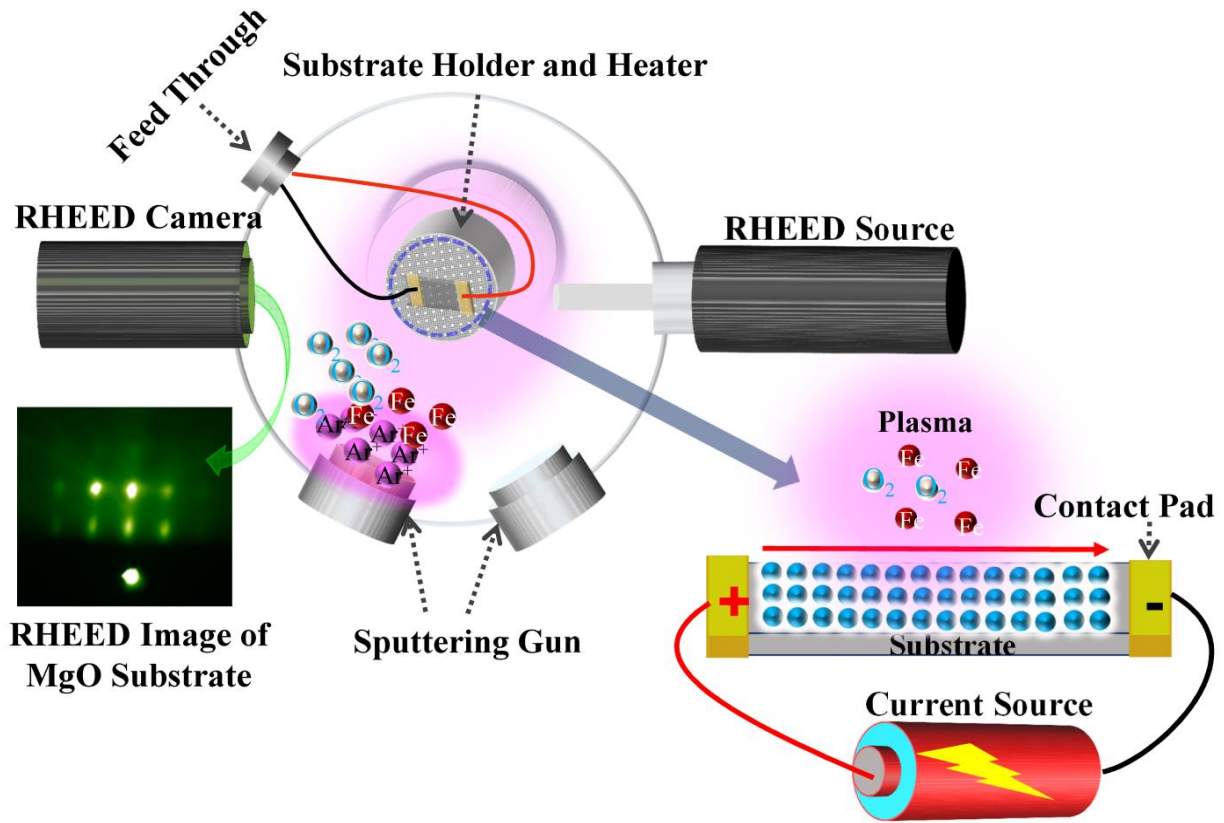

**Figure 2.**

**Figure 2** $\theta/2\theta$ patterns for the ASD, ED1 and ED2 epitaxial Fe$_3$O$_4$(100)/MgO(100) thin films.

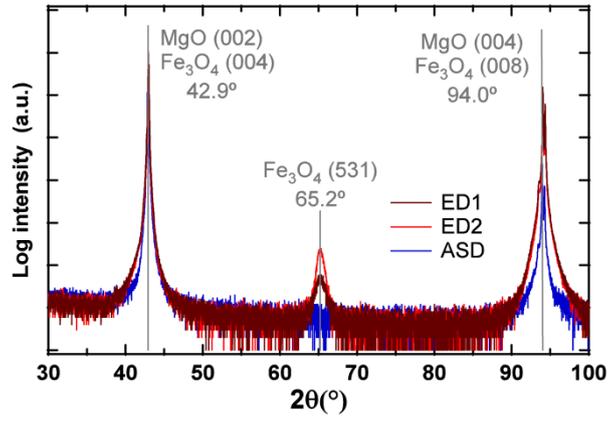

**Figure 3.**

**Figure 3** Pole figures for the different samples; (a) ASD, (b) ED1, and (c) ED2. Peaks corresponding to (001) oriented $Fe_3O_4$ crystals visible at tilt angles ($\psi$) 25.2° and 72.5°; peaks from MgO(111) reflections, corresponding to the (001) oriented MgO substrate, visible at $\psi$ = 54.7°.

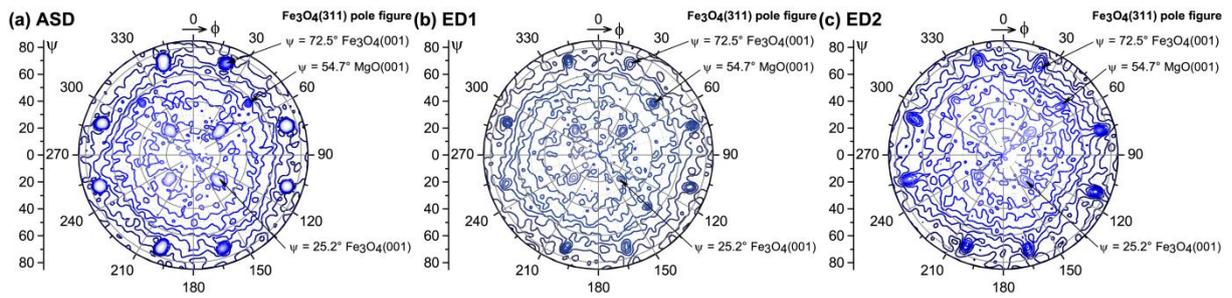

**Figure 4.**

**Figure 4** Dark- and bright-field TEM images of the ASD and ED2 samples. (a) Dark-field image ([001]-zone axis) of the ASD sample. The image was formed using 220 reflection (marked by the diffraction pattern in the inset) to visualize the anti-phase boundaries in the single crystalline $Fe_3O_4$ film. The dark contrast bands at the image border are due to a slight bending of the free-standing film. (b) Bright-field image of the ED2 sample. (c) Diffraction pattern corresponding to the area in white circle in (b). (d) Dark-field image, taken using the 220 reflection in the diffraction in (c). (e) Detail of the grain structure in the ED2 film.

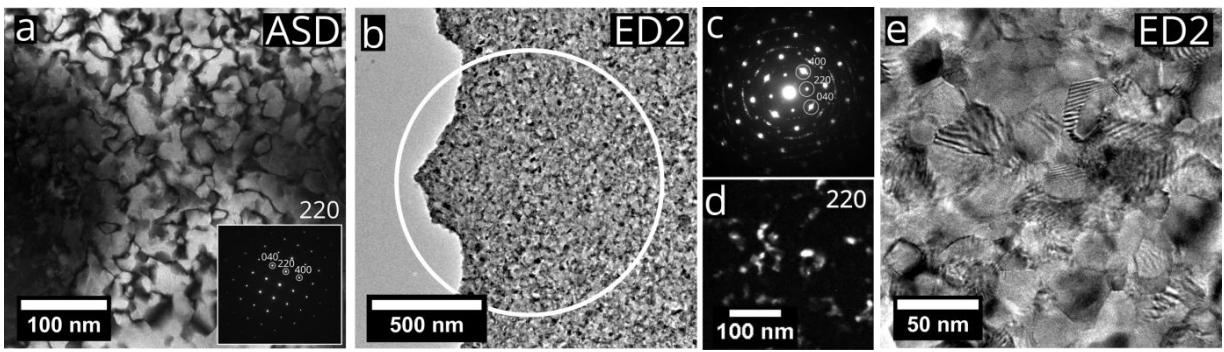

**Figure 5.**

**Figure 5** Raman symmetric stretch vibration $A_{1g}$ mode spectra for (a) ASD, (b) ED1, and (c) ED2 films, and antisymmetric and symmetric bending $T_{2g}(3)$ mode spectra for (d) ASD, (e) ED1, and (f) ED2 films. Red lines are Lorentz fits of the data to extract the line shape parameters.

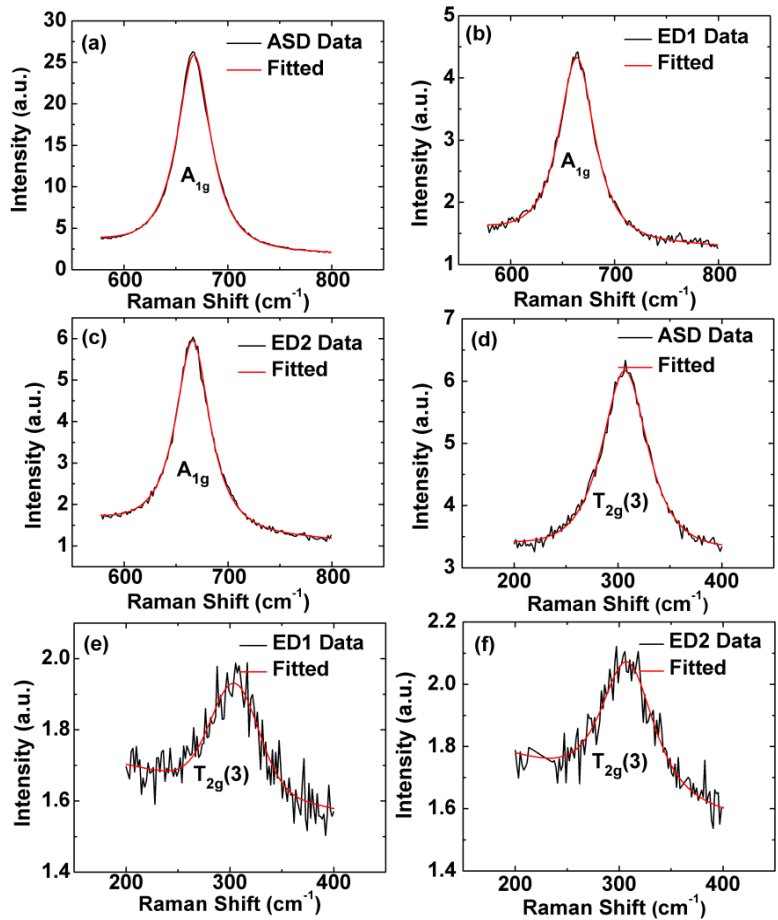

**Figure 6.**

**Figure 6** In-plane magnetization versus field recorded at different temperatures for (a) ASD, (b) ED1, and (c) ED2. Insets show $M_s = M(400 \text{ mT})$ versus temperature for the different samples. It should be noted that a field of 400 mT is not enough to saturate the magnetization for the ASD sample, which becomes particularly evident at temperatures below the Verwey transition. (d) In-plane $M/M_s$ versus field recorded at 300 K for the different samples. (e) Magnetization versus field recorded along 0° (MgO[001]), 45° (MgO[011]), and 90° (MgO[010]) ED1 film azimuths. (f) Normalized in-plane and out-of-plane magnetization versus field for the ED1 and ASD films.

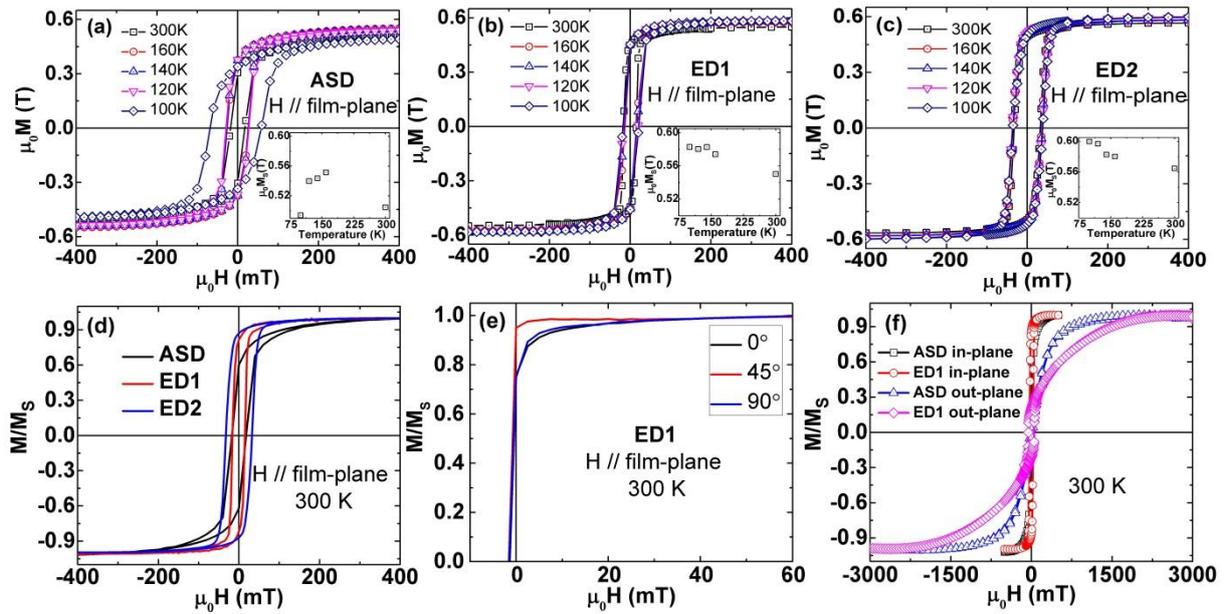

**Figure 7.**

**Figure 7** Normalized field-cooled magnetization $M/M_V$ versus temperature to probe the Verwey transition of the ASD and ED2 samples. $\mu_0 H = 10$ mT and $M_V$ is the magnetization at the Verwey transition.

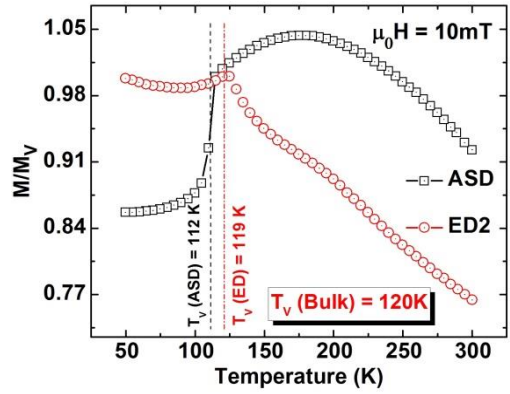